\documentclass[journal=mamobx,manuscript=letter]{achemso}
\usepackage{graphicx}
\usepackage{epstopdf}
\usepackage{comment}
\usepackage{textcomp}
\usepackage{gensymb}
\usepackage[T1]{fontenc}
\usepackage{amsmath,amssymb} 
\usepackage{soul}               
\usepackage{color,colordvi}
\usepackage{makecell}
\mciteErrorOnUnknownfalse

\AppendGraphicsExtensions{.tif}

\date{\today}

\author{Andrea Ninarello$^{1,2}$}
\author{J\'{e}r\^{o}me J. Crassous$^{3,4}$} \email{crassous@pc.rwth-aachen.de}
\author{Divya Paloli$^4$}
\author{Fabrizio Camerin$^{1,5}$}
\author{Nicoletta Gnan$^{1,2}$}
\author{Lorenzo Rovigatti$^{2,1}$}
\author{Peter Schurtenberger$^4$}
\author{Emanuela Zaccarelli$^{1,2}$} \email{emanuela.zaccarelli@cnr.it}
\affiliation{$^1$ CNR-ISC Uos Sapienza, Piazzale A. Moro 2, IT-00185 Roma, Italy}
\affiliation{$^2$ Department of Physics, {\em Sapienza} Universit\`a di Roma, Piazzale A. Moro 2, IT-00185 Roma, Italy}
\affiliation{$^3$ Institute of Physical Chemistry, RWTH Aachen University, Landoltweg 2, DE-52074 Aachen, Germany}
\affiliation{$^4$ Physical Chemistry, Department of Chemistry, Lund University, Naturvetarv\"agen 14, SE-22100 Lund, Sweden}
\affiliation{$^5$ Department of Basic and Applied Sciences for Engineering, {\em Sapienza} Universit\`a di Roma, via A. Scarpa 14, IT-00161 Roma, Italy}

\title{Modelling microgels with controlled structure across the volume phase transition}

\begin{document}

\begin{abstract}
Thermoresponsive microgels are soft colloids that find widespread use as model systems for soft matter physics. Their complex internal architecture, made of a disordered and heterogeneous polymer network, has been so far a major challenge for computer simulations.
In this work we put forward a coarse-grained model of microgels whose structural properties are in quantitative agreement with results obtained with small-angle X-ray scattering experiments across a wide range of temperatures, encompassing the volume phase transition. These results bridge the gap between experiments and simulations of individual microgel particles, paving the way to theoretically address open questions about their bulk properties with unprecedented nano and microscale resolution.
\end{abstract}

\maketitle

\section{Introduction}

As many fields of physics, colloidal science has profoundly benefited from the interplay between experiments and computer simulations. In the last decades, the epitome of colloids in suspension was represented by Poly(methyl methacrylate) (PMMA) particles in experiments~\cite{Bartlett1994} and by their numerical analogs, the hard sphere model~\cite{Frenkel}.
However, in more recent years, colloids with internal degrees of freedom and tunable softness have been progressively replacing hard spheres in experimental studies~\cite{vlassopoulos2014tunable}. Among these, microgels, i.e. spherical particles made of a crosslinked polymer network, with a radius ranging from 20~nm to 500~nm and typically synthesized with radical emulsion or precipitation polymerization, have become particularly popular~\cite{MicrogelBook} for two main reasons: (i) a relatively easy preparation protocol yielding quite monodisperse particles and (ii) the possibility of finely tuning the particle properties by changing the chemical composition of the constituent polymers. For example, in the case of thermoresponsive polymers such as Poly(N-isopropylacrylamide) (PNIPAM), microgels undergo a so-called Volume Phase Transition (VPT) at a temperature $T_{VPT}\sim 32^{\circ}$C across which they change from a swollen state at low temperature to a compact state at high temperatures. Due to their high versatility, microgels have been used to address numerous open problems in condensed matter science, including the fabrication of responsive colloidosome~\cite{Kim2007}, the nucleation of squeezable particles~\cite{Iyer2009,Frenkel2009,Scotti2016}, the premelting within crystalline states~\cite{Alsayed2005}, and the glass and jamming transition of soft colloids~\cite{Mattsson2009,Paloli2013,Philippe2018,Zhang2009}.

The downside of using PNIPAM microgels is often an incomplete control on the internal particle topology~\cite{Lyon2012}, usually comprising
a dense core and an outer corona, which is rather heterogeneous and mainly composed of long chains and few crosslinkers~\cite{MicrogelBook}. One of the most successful descriptions of the internal topology of microgels is the so-called fuzzy sphere model, in which a strictly homogeneous core is surrounded by a loose corona~\cite{Stieger2004}. The use of super-resolution microscopy~\cite{Conley2016, Gelissen2016,Conley2017} recently allowed the real space visualization of the internal density profiles of the microgels, revealing that the core is not really homogeneous, being denser in its inner part and progressively rarifying towards the corona~\cite{Bergmann2018, Siemes2018}. 

While extensive characterization of the microgel internal structure has been provided experimentally, modelling and simulations have so far lagged behind. Indeed, the description of the inherently multiscale nature of these particles, from the polymeric constituents up to the colloidal scale, is a demanding task,
even augmented by the disordered and heterogeneous structure of the network.
Several numerical works modelled microgels using a polymeric crystalline lattice~\cite{Escobedo1999, Jha2011, Kobayashi2014,Ahualli2017},
while only recently the investigation of disordered crosslinked networks {\it in silico} has been reported~\cite{Nikolov2017,Gnan2017,Moreno2018}.
However, all the methods proposed so far were unable to finely control the internal density distribution of the network and, consequently, to reproduce the properties of the experimentally available microgels in a truly quantitative fashion.

In this article, we put forward a novel numerical methodology where microgels with desired internal density profiles are generated. Building on the assembly protocol proposed in Ref.~\cite{Gnan2017}, we introduce a designing force on the crosslinkers that is able to tune the core-corona architecture independently of the system size. By carefully adjusting the force field and intensity, we obtain individual microgel particles that quantitatively reproduce the experimentally measured form factors and swelling behavior across the VPT. We also quantify the effect of coarse-graining on the structure of the \textit{in silico} microgels by performing our investigation as a function of the simulated system size. Our results move numerical and experimental investigation of microgels at the single particle level closer to each others, providing a realistic description of these soft colloids at all relevant scales and paving the way to a deeper understanding of their collective behaviour.

\section{Models and Methods}

\subsection{Numerical methods}
\begin{figure}[!htb]
\includegraphics[width=0.4\textwidth]{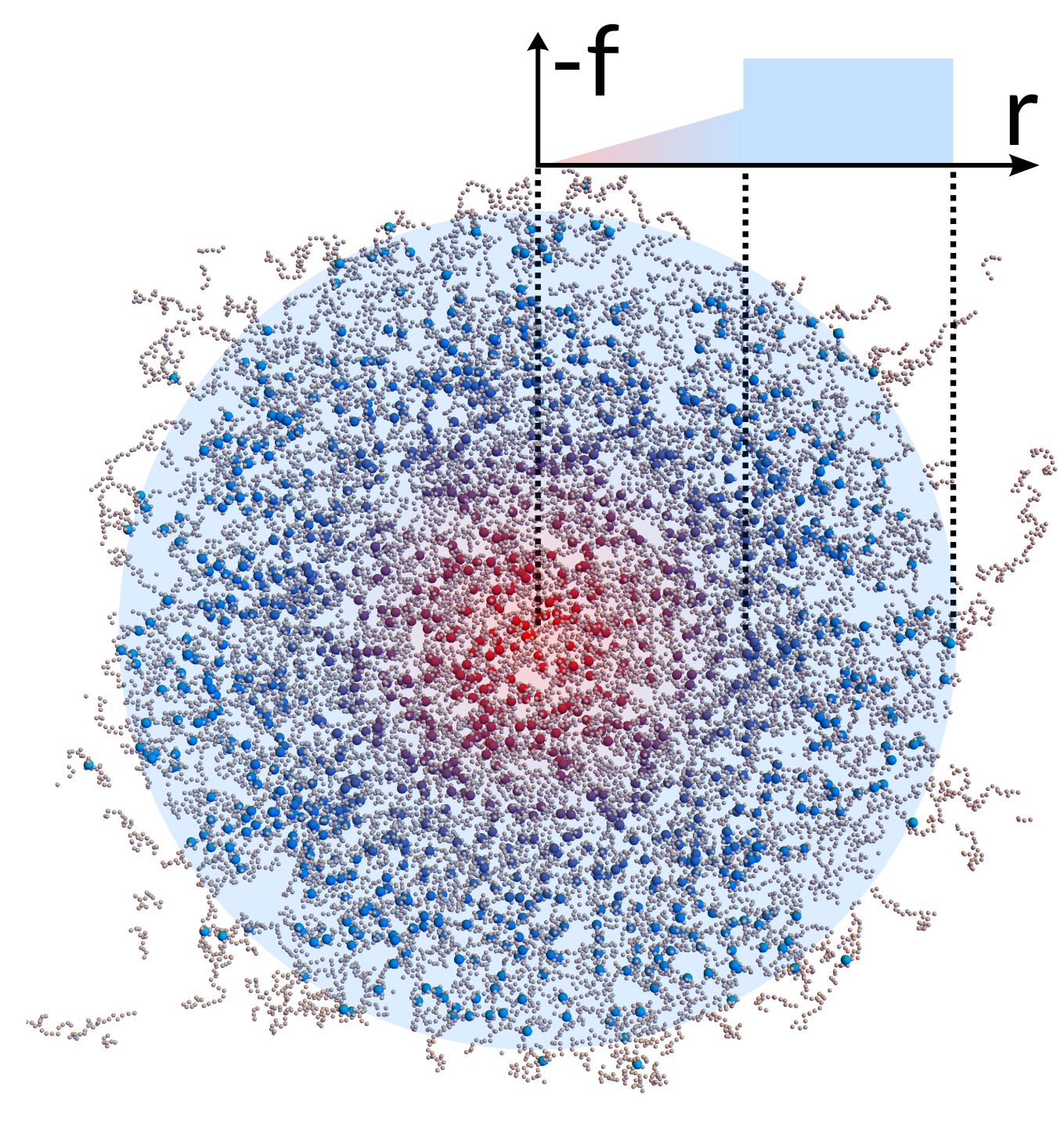}
\caption{\label{sketch_force} Snapshot of a $N\approx 336000$ microgel slice of width $20\sigma$. Monomers are represented in grey, while
crosslinkers (magnified in size with respect to monomers to improve visualization) are coloured from red in the center to blue in the corona, following the shape of the designing force of magnitude $f$, which is illustrated in the top right corner. The use of the force correctly imposes an inhomogeneous profile to the microgel, with a larger concentration of crosslinkers in the core region. At the microgel boundary, few chains are disconnected since in this representation parts of the corona are outside the field of view. The attractive force on the crosslinkers used during the assembly is made of two contributions: (i) an elastic force of spring constant $k$ acting from the center of the cavity up to a distance $\frac{Z}{2}$ and (ii) a gravity-like force of strength $g$ being at work between $\frac{Z}{2}$ and $Z$. Extensive details about this choice are provided in the Results section.
}
\label{fig1}
\end{figure}

We generate microgels exploiting the method put forward in Ref.~\cite{Gnan2017} through which it is possible to obtain fully-bonded, disordered polymer networks. To this aim, we simulate the self-assembly of a binary mixture of patchy particles covered by two and four attractive patches which represent monomers and crosslinkers, respectively. During the assembly stage the simulations are performed in a spherical cavity of radius $Z$. The total number of particles is $N$ and the crosslinker concentration is fixed to $c=5\%$.
Patchy particles interact via the sum of a Weeks-Chandler-Andersen (WCA) repulsion~\cite{Weeks1971} and
a previously employed attractive patchy potential~\cite{Rovigatti2018}. In order to accelerate the network formation, we employ a bond-swapping algorithm~\cite{Sciortino2017}.
We perform NVT molecular dynamics simulations of $N=5000, 42000, 336000$ monomers of unit mass $m$ confined in spherical cavities respectively of diameter
$Z=25, 50, 100$ in units of bead size $\sigma$ in order to maintain the final number density of the microgel roughly constant.
In order to control the monomer internal density distribution, and in particular the width of the corona, we here extend the previous method of  Ref.~\cite{Gnan2017} by introducing a designing force  $f(r)$, which pulls crosslinkers towards the center of the cavity.
Detailed information about the choice of the force is provided in the Results section. An illustration of the assembled microgel and of the designing force is reported in Fig.~\ref{fig1}.

Once the network is built, we make the bonds between monomers permanent, by replacing the patchy attraction with a Finite Extensible Nonlinear Elastic (FENE)
potential for bonded monomers, leaving unaltered  the WCA repulsion. This is the well-known Kremer-Grest potential widely employed to investigate numerically polymeric systems~\cite{Kremer1990}.  
To take into account the effect of temperature, we consider an additional solvophobic potential~\cite{Soddemann,LoVerso2015,Gnan2017}, which mimics the reduced affinity of the
monomer to the solvent with increasing $T$, of the following form:
\begin{equation}
  \label{va_force}
  V_\alpha(r)=
  \begin{cases}
    -\epsilon \alpha \ \ \        &\text{if}\ r\leq 2^{1/6}\sigma,\\
    \frac{1}{2}\alpha \epsilon  \big[ \cos(\gamma(r/\sigma)^2+\beta)-1 \big]  \ \ \     &\text{if}\ 2^{1/6}\sigma<r<R_0\sigma,\\
    0 \ \ \ &\text{if}\ r>R_0\sigma.
  \end{cases} 
\end{equation}
Here $\gamma=(\pi(2.25-2^{1/3}))^{-1}$, $\beta = 2\pi-2.25\gamma$ and $\epsilon$ is the unit of energy.  The $V_\alpha$ potential is  modulated by the solvophobic
parameter $\alpha$ which controls the strength of the monomer-monomer attractive interactions. Thus, $\alpha=0$ corresponds to the case where there is no
attraction and the microgel is maximally swollen.\\ 
We perform Nose-Hoover molecular dynamics runs with the LAMMPS package~\cite{LAMMPS} with a time step $\delta t =0.002$ at fixed temperature $k_B T=1$ where $k_B$ is
the Boltzmann constant. Single microgels are at first equilibrated for $1\times10^6$ time steps until their radius of gyration $R_g$ reaches a constant value. After
equilibration, a production run is performed for up to $1\times10^7$ time steps, saving configurations every $5\times10^5$ steps. To improve statistics, for all system sizes,
we average results over ten independent microgel topologies. We calculate the form factor of the microgels defined as
\begin{equation}
  P(q)=\frac{1}{N}\sum_{ij}\langle\exp(-i\vec{q}\cdot\vec{r}_{ij})\rangle,
  \label{eq:Pq}
\end{equation}
where the angular brackets indicate an average over independent configurations, $q$ is the wavevector and $r_{ij}$ is the distance between monomers $i$ and $j$.

From simulations, we also readily calculate the radial density profile $\rho(r)$ as a function of the distance $r$ from the center of mass of the microgel. The two observables $\rho(r)$ and $P(q)$ allow us to compare the structure of the {\it in silico} microgels with experiments both in real and in reciprocal space.

One of the most successful models to describe microgel density distribution is the widely employed fuzzy sphere model~\cite{Stieger2004},
where a microgel is considered as a sphere of radius $R'$ with a homogeneous core surrounded by a fuzzy corona.
Recent results from super-resolution microscopy have shown a slightly inhomogeneous core~\cite{Bergmann2018} and put forward a generalization of the fuzzy sphere model with the addition of a linear dependence of the density inside the core.

In real space, this so-called extended fuzzy sphere model is represented by an error function multiplied by a linear term:
\begin{equation}
  \rho(r) \propto \text{Erfc} \left[\frac{r-R'}{\sqrt{2}\sigma_{\rm surf}}\right] (1-sr),
  \label{eq:rho}
\end{equation}
where $R'$ corresponds to the radius at which the profile has decreased to half the core density, $\sigma_{\rm surf}$ quantifies the width of the corona and $s$ is the slope
of the linear decay. We notice that for $s=0$ the standard fuzzy sphere model is recovered.

Equation ~\ref{eq:rho} can be written in Fourier space as:
\begin{align}
  \label{fuzzyX}
  P(q) \propto &\left\{ \left[ \frac{3(\sin(qR)-qR\cos(qR))}{(qR)^3} \right.\right.\nonumber \\
  +& \left.\left. s\left(\frac{\cos(qR)}{q^2R}-\frac{2\sin(qR)}{q^3R^2}-\frac{\cos(qR)-1}{q^4R^3}\right)\right] \times  \exp\left[-\frac{(\sigma q)^2}{2}\right]\right\}^2.
\end{align}
We will adopt the extended fuzzy sphere model in the following to describe experimental form factors and to extract the associated density profiles, that will be then compared to those directly calculated from simulations.

\subsection{Experimental details}

PNIPAM microgels were synthesized by surfactant free radical polymerization as described in former studies~\cite{Paloli2013,Mohanty2014}.
NIPAM ($2$~g) as monomers,  N,N'-methylenebisacrylamide (BIS, $0.136$~g) as cross-linker, and methacryloxyethyl thiocarbomoyl rhodamine B
($2$~mg dissolved in $87.8$~mL of water) as the dye were polymerized by precipitation polymerization. The reaction was initiated by dropwised
addition of sodium dodecylpersulfate initiator ($0.01$~g in $10$~mL of water) at $80~\degree {\rm C}$ and run for $4$~h under constant stirring
at $300$~rpm and nitrogen purging. The reaction mixture was passed through glass wool in order to remove particulate matter. The dispersions
were purified by repeated centrifugation/redispersion cycles against an aqueous $10^{-3} $ M potassium chloride (KCl) solution. The different suspensions
were further obtained by dilution of the stock suspension with the aqueous KCl solution.

Experiments were performed at the Swiss Light Source (SLS, Paul Scherrer Institute) at the cSAXS instrument. A X-ray beam with an energy of $11.2$~keV
was used, corresponding to a wavelength $\lambda =0.111$~nm. The $q$-scale was calibrated by a measurement of silver behenate. No absolute calibration
was done for the X-ray data. The sample consists of a $1$~wt$\%$ microgel dispersion containing $10^{-3}$ M KCl enclosed in a $1$~mm diameter sealed
quartz capillary (Hilgenberg GmbH, Malsfeld, Germany) placed in a homemade thermostated aluminum sample holder ensuring a temperature control with an
accuracy of $0.2~\degree {\rm C}$. At least $30$ 2D images were taken, azimuthally integrated, transmission and background corrected, and averaged
according to established procedures provided by PSI.

Experiments were carried out using a light scattering goniometer instrument from LS Instruments equipped with a HeNe laser light source with a
wavelength $\lambda = 632.8$~nm and a maximum power of $35$~mW. The sample was filled into cylindrical NMR tube of a diameter of $5$~mm and placed
in the temperature controlled index matching bath ($\pm 0.1~\degree {\rm C}$). The scattered light was detected by two APD detectors and processed by
a Flex correlator in cross-correlation  configuration. A modulation unit was employed as recently described by Block et al.~\cite{Block2010}.
All the measurements were performed on an aqueous $0.01$~wt$\%$ suspension containing $10^{-3}$ M KCl. The scattering angle $\theta$ was varied from
30$^\circ$ to 50$^\circ$ every 5$^\circ$. The initial decay rate $\Gamma_0$ was derived from a first cumulant analysis of the normalized field autocorrelation
function. The diffusion coefficient D$_0$ was estimated from the $q^2$-dependence $\Gamma_0 = D_0q^2$, and the hydrodynamic radius $R_H$ obtained via the
Stokes-Einstein relation $D_0 = k_BT/(6\pi\eta_s R_H)$, where $k_B$, $\eta_s$, and $T$ are the Boltzmann constant, solvent viscosity, and absolute temperature, respectively.

\section{Results}
\subsection{The choice of the designing force}

One of the main aims of this study is to set up a protocol being able to finely control the radial density distribution of the microgel.
In this section, we discuss how to implement this feature using a designing force during the self-assembly of the patchy particles mimicking monomers (bivalent) and crosslinkers (tetravalent) in a spherical cavity.

\begin{figure}[!htb]
  \centering
  \includegraphics[width=0.98\textwidth]{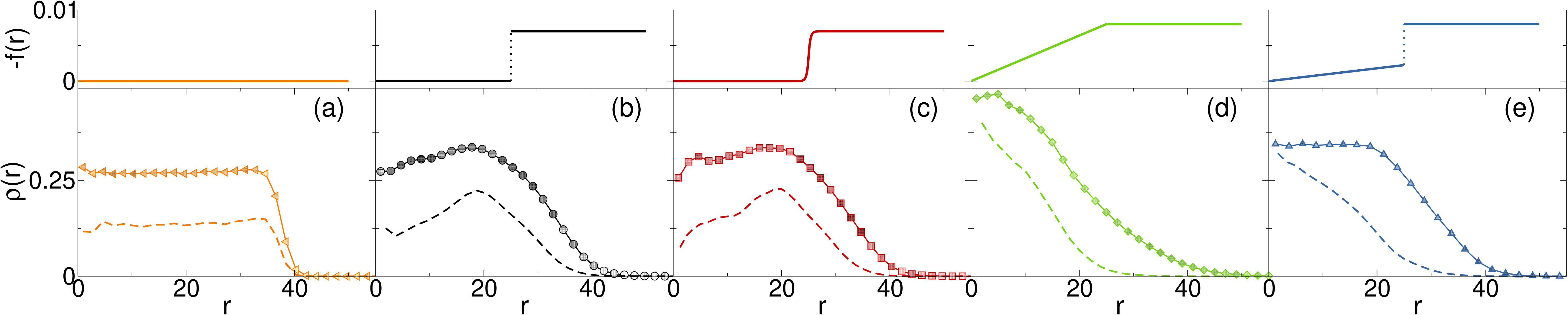}
  \caption{ \label{forces} Different types of forces acting on crosslinkers (top panel) and corresponding density profiles for all particles (symbols) and for crosslinkers only (dashed lines). In the five panels different inward forces, acting only on crosslinkers. are considered: (a) No force, (b) a force as in Eq.~\ref{forces_first} with $g=8 \times 10^{-3}$ and $k=0$; (c) a force as in Eq.~\eqref{forces_second} with $m=7\times 10^{-3}$ and $t=0.3$; (d) a force as in Eq.~\ref{forces_first} with $g=8\times10^{-3}$ and $k=\frac{2g_1}{Z}=3.2\times 10^{-4}$; (e) a force as in Eq.~\ref{forces_first} with $g=8\times10^{-3}$ and $k=4.5\times 10^{-5}$. In all cases, the integral of $\rho(r)$ is normalized to a constant value $\int \rho(r) dr =c$ with $c=10,5$ for all particles and crosslinkers, respectively, to improve visualisation. Data are averaged over four independent realizations obtained with the numerical protocol described in the Methods section. Case (e) is the one finally adopted to compare with experiments in the following sections.}
\end{figure}

In this work we specifically target the reproduction of the topology of PNIPAM microgels synthesized using free radical precipitation polymerization. For these particles, the core slowly rarefies from the center towards the corona, resulting in linearly decreasing density profiles, as observed through super-resolution microscopy~\cite{Bergmann2018}. Also the corona should be reproduced with the correct width and shape.  The fact that microgels have a denser core is the result of a faster consumption of the
crosslinker with respect to NIPAM~\cite{Stieger2004} during the polymerization process, resulting in a larger crosslinker concentration within the core. To obtain such an inhomogenous crosslinker distribution within the microgel, we apply a force acting on crosslinkers only.  Indeed, if the force is applied on all the monomers, the resulting density profiles is much more homogeneous than in experiments.

However, the exact shape that the force should assume is not {\it a priori} obvious. 
In order to obtain the desired density profile, we have tested different functional forms of the force and compared the results with the unperturbed case, i.e. the assembly in the absence of a force that was adopted in Ref.~\cite{Gnan2017}.
In all cases, the assembly is carried out by fixing the total number of particles to $N=42000$ with a fraction of crosslinkers equal to $5\%$. 
We confine the system in a spherical cavity of radius $Z$, which determines the number density and the size of the final microgel. Using too small or too large values of $Z$ gives rise to microgels that are either too compact or too fluffy, very far from the realistic core-corona structure. We thus select the intermediate value of $Z=50\sigma$, which correspond to a number density $\rho \sim 0.08$, that provides the best conditions to reproduce experiments with the additional force on the crosslinkers. All the configurations are realised using the protocol described in the Models and Methods section.

In Fig.~\ref{forces} we report an illustration of different choices of the designing force as a function of the distance from the center (top panels) and the associated density profiles (bottom panels) for all the monomers (symbols) and for crosslinkers only (dashed lines).
In the absence of a designing force, shown in Fig~\ref{forces}(a), we find that the microgel is made of a homogeneous core and of a rapidly decaying corona. This is reflected by the flat density profile of the crosslinkers. The situation gets worse when we increase the microgel size: since the decay of the corona happens only at the microgel surface, the increase of the volume/surface ratio gives rise to an unrealistically thin corona. Ideally, instead, we would like to mantain the same ratio of the size of the corona with respect to the width of the core  (corona-core ratio) when we vary the microgel size, in order to have a valid protocol that is applicable to any $N$. 
Thus, we need to control the width of the corona and to this aim, we apply an inward force with spherical symmetry inside the cavity.

We have considered two types of  forces. The first type is described by the following expression:
\begin{equation}
  \label{forces_first}
  \vec{f}_1(r) =
  \begin{cases}
    -k r\hat{r}\ \ \        &\text{if}\ 0<r\leq C\\
    -g \hat{r}\ \ \     &\text{if}\ C <r<Z,
  \end{cases}
\end{equation}
where $\hat{r}$ is a versor pointing outward. Here an elastic force with a coefficient $k$ acts from the center up to the half radius of the cavity and a force of
constant $g$ is present  for larger distances. We choose $C=\frac{Z}{2}$ as the point where the force changes type in order to reproduce a core corona structure
for the microgel.  We verified that the shape of the resulting microgel is nearly the same for values of this point up to $3Z/5$. The second type of force smooths
out the discontinuity at $Z/2$, increasing continuously from the center to the cavity boundary:

\begin{equation}
  \label{forces_second}
  \vec{f}_2(r) = \begin{cases}
    -\left[\frac{m}{2}\exp\left(\frac{r- C}{t}\right)\right]\hat{r}   & \text{if}\ 0<r\leq C\\
    -\left[m-\frac{m}{2}\exp\left(-\frac{r-C}{t}\right)\right]\hat{r} & \text{if}\ C<r<Z.
  \end{cases} 
\end{equation}

Here $m,t$ determine the strength and the smoothness of the force, respectively.
We use again $C=\frac{Z}{2}$.

Initially, we consider a force $f$ of type $f_1$ with constant $g=8 \times 10^{-3}$ and $k=0$, shown in Fig.~\ref{forces}(b). One can observe that, although
the corona becomes larger, the core is sparser for small $r$ and denser close to the corona. This entails the emergence of a peak at $r\lesssim Z/2$ showing that
crosslinkers tend to accumulate around this particular distance and their number decreases towards the center of core, which is not compatible with experimental findings for the class of microgels used in this study.
Since the presence of a peak could be due to the discontinuity of $f$ at $Z/2$, we have also employed a smooth force of type $f_2$
by Eq.~\eqref{forces_second}. However, in this case, independently of the choice of the force parameters,  the peak is not removed. The choice $m=7\times 10^{-3}$ and $t=0.3$
provides a density profile very similar to the previous one (see Fig.~\ref{forces}(c)) for both monomers and crosslinkers. 
One can then conclude that the additional peak is not given by the discontinuity itself but it is a consequence of
the weakness or absence of the force in the region $0<r<Z/2$.
Therefore, our next attempt is to maintain the corona shape of the previous examples and get rid of the peak.
To this aim, we again employ a force of type $f_1$ with $g=8\times10^{-3}$ and $k=\frac{2g_1}{Z}=3.2\times 10^{-4}$. The use of $k\neq 0$
corresponds to the application of an elastic force in the inner half region, Eq.~\eqref{forces_first} which is continuous at $Z/2$.
Furthermore, we employ the same value of $g$ as before in order to keep unchanged the shape of the corona.
The resulting density profile is reported in Fig.~\ref{forces}(d).
In this case, we notice that the density distribution inside the microgel is strongly altered, with
a continuously decreasing density from the center to the cavity boundary. The absence of a core is totally different 
from experimental observations.

We infer that this effect is a consequence of the intensity of the force for $r<Z/2$, and therefore we decide to decrease the
spring constant of the force as sketched in Fig.~\ref{forces}(e), resulting in a discontinuity at $Z/2$. 
Using the value $k=4.5\times 10^{-5}$, we find a density distribution in the core in agreement with the experiments, while preserving
the right shape of the corona. Interestingly, in this case, the crosslinker profile is continuously decreasing from the center of the microgel and does not reflect the total profile of all the monomers. This is the choice that we adopt in the following to reproduce the experimental results for all studied system sizes.

\subsection{Size effects}
It is important to investigate the robustness of our results with respect to system size. Of course, the use of a large number of monomers provides a remarkable improvement in the quantitative comparison with experiments, at the cost of a huge increase of the required computational resources. Hence, we need to identify the optimal system size to use in computer simulations in order to be able to tackle a specific problem. To this aim, this section is dedicated to the comparison of the structure the three studied system sizes assembled in presence of the force and of a large microgel in which the force has been set to zero during the assembly (unperturbed).

\begin{figure}[!htb]
  \minipage{0.32\textwidth}
  \includegraphics[width=\linewidth]{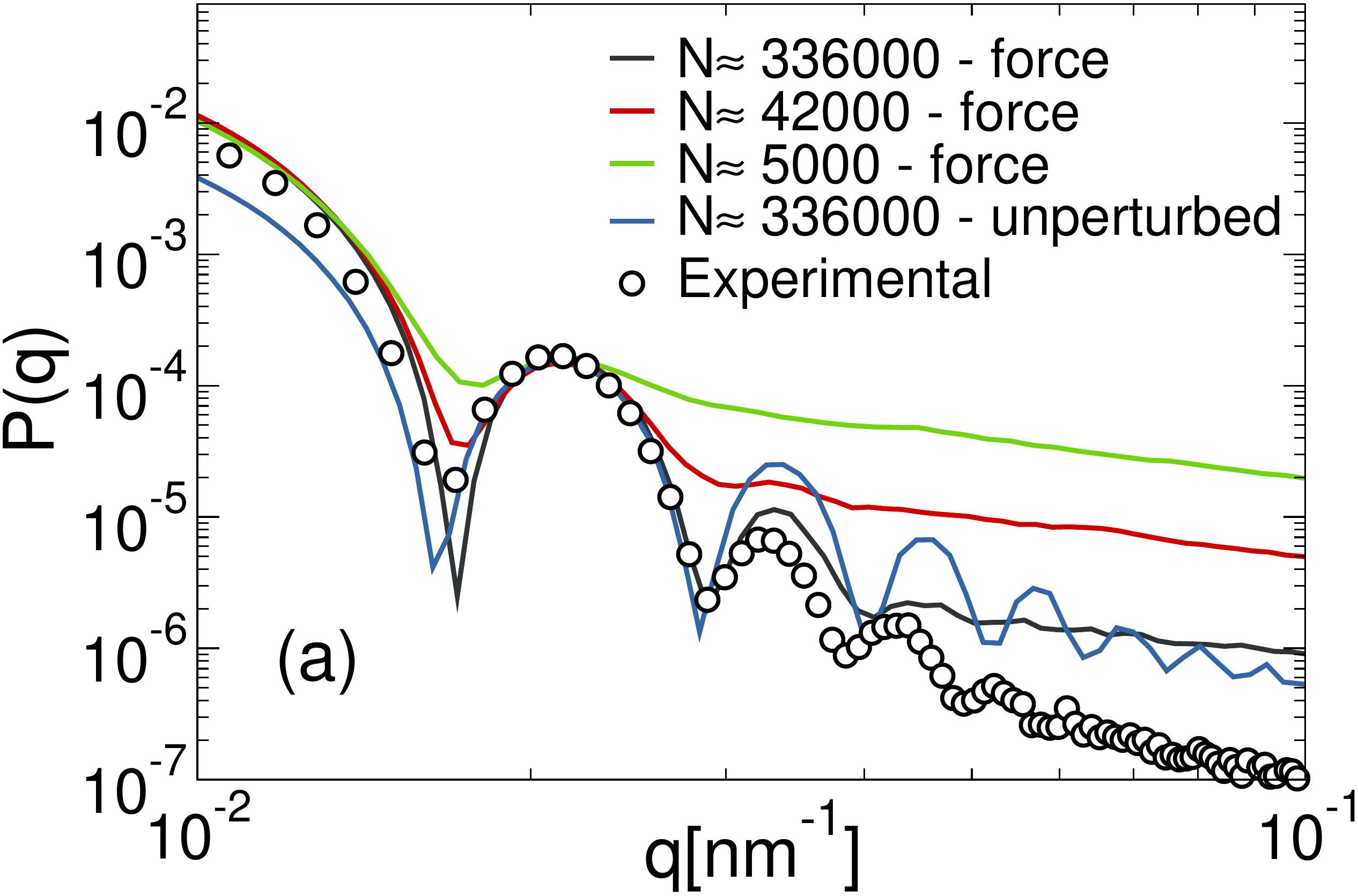}
  \endminipage\hfill
   \minipage{0.313\textwidth}
   \includegraphics[width=\linewidth]{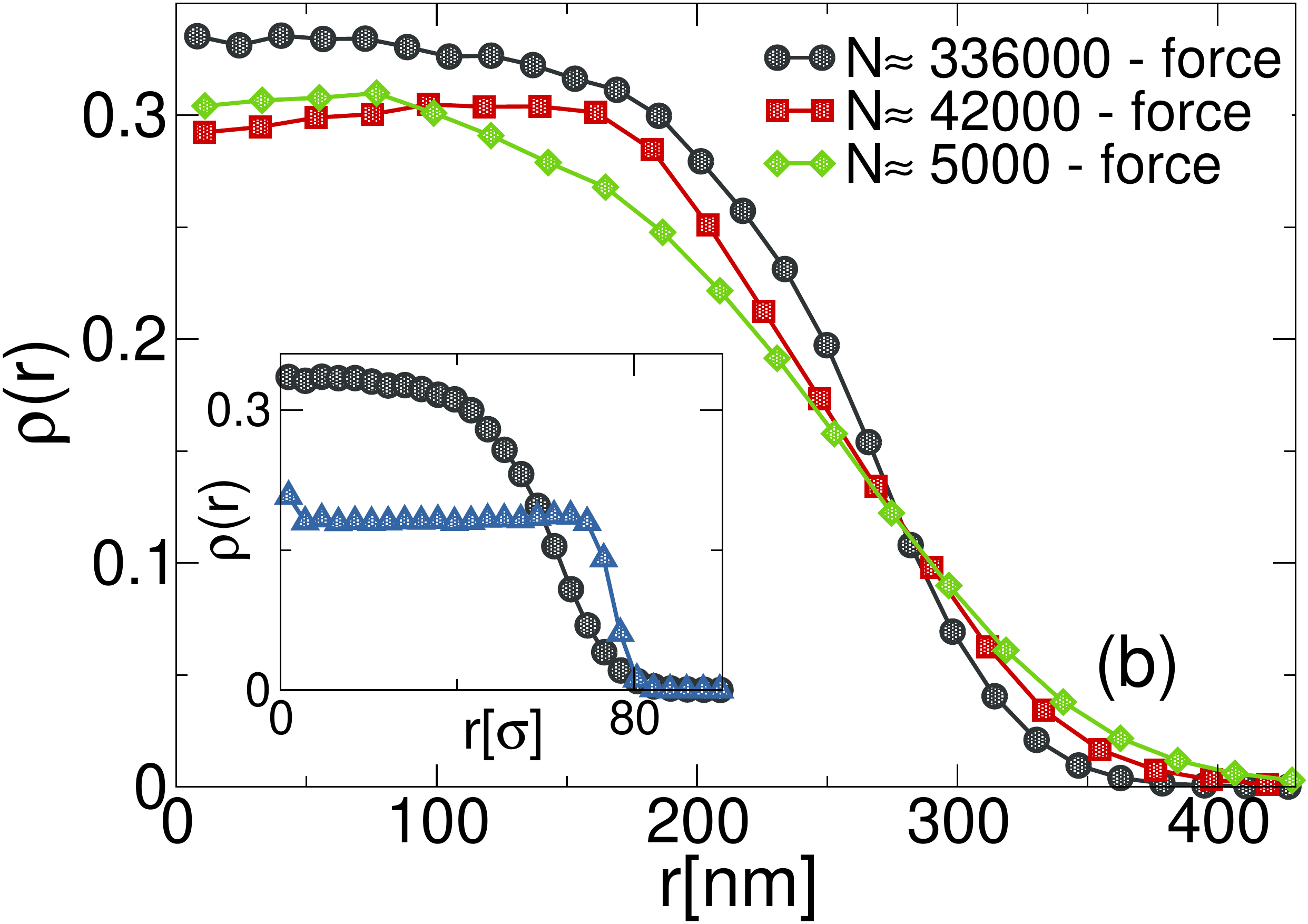}
  \endminipage\hfill
   \minipage{0.32\textwidth}
  \includegraphics[width=\linewidth]{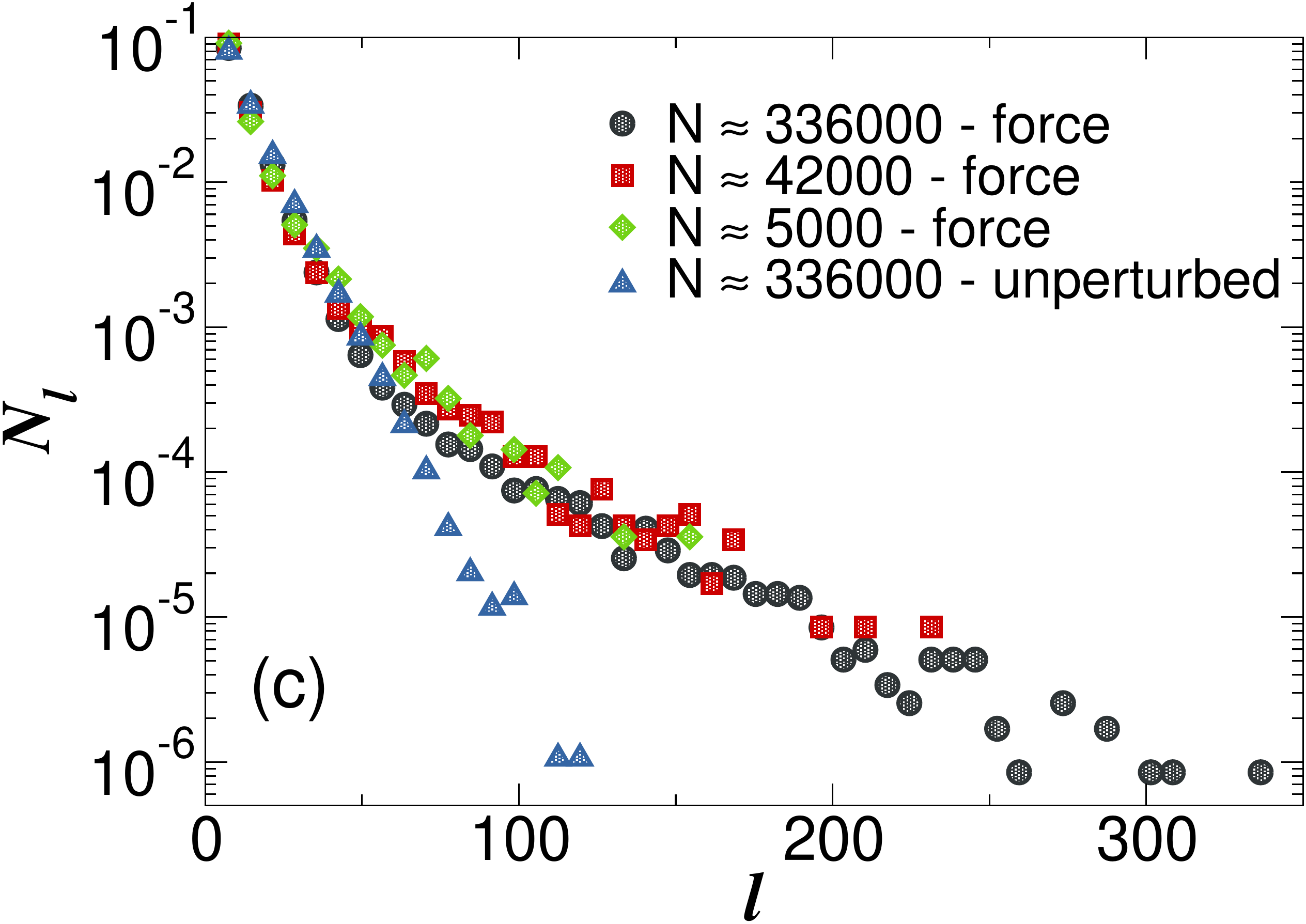}
  \endminipage\hfill
  \includegraphics[width=\linewidth]{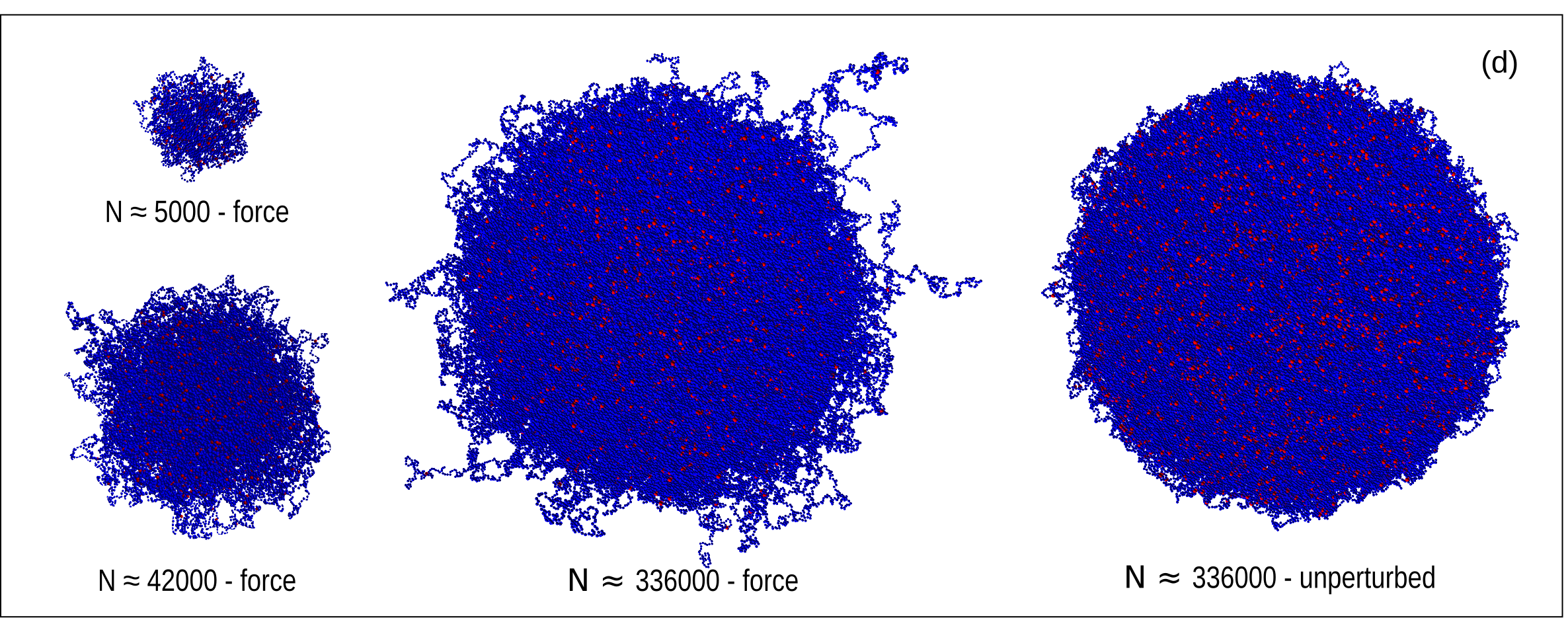}
 \caption{  \label{size_effects}
   Size effects on the structural properties of the microgels for three system sizes obtained with the same designing force and in the unperturbed case:
   (a)  numerical form factors $P(q)$ in the swollen state ($\alpha=0$) . The data are compared with experimental measurements for $T=15.6\degree$ C (black circles) through the rescaling factors $\gamma_{336000}=0.233$, $\gamma_{42000}=0.124$, $\gamma_{5000}=0.0580$ and $\gamma^{unpertubed}_{5000}=0.274$;  (b) density profiles of the three simulated microgels, scaled on the $x$-axis by $1/\gamma$ for the corresponding size. Inset: density profiles for $N\approx 336000$ systems in units of $\sigma$;  (c) chain length distribution $N_l$;
(d) snapshots with monomers represented in blue and crosslinkers in red.
}
\end{figure}

The effect of the microgel size is evident in the behavior of the form factors $P(q)$, reported in Fig.~\ref{size_effects}(a), for the swollen state  ($\alpha=0$). The numerical data for different $N$ are compared to the experimental form factor for the lowest measured temperature ($T=15.6\degree C$), that we set to be the maximally swollen case in our model. In order to perform the comparison, we match the position of the first peak, $q^*_{\rm sim}$, of the numerical $P(q)$ onto that of the experiments, $q^*_{\rm exp}$. This procedure defines the scaling factor $\gamma=q^*_{\rm exp}/q^*_{\rm sim}$ that allows to convert numerical units into real ones.

We observe that the first peak of $P(q)$ for the smallest system ($N\approx 5000$) is just barely visible, whereas it becomes better defined by increasing the microgel radius by a factor of $\sim 2$ ($N\approx 42000$), with the simultaneous appearance of a second peak. Finally, the largest system tested ($N\approx 336000$), corresponding to a further increase by a factor of $\sim 2$ in radius, reproduces quite well three out of the four peaks observed in the experimental curve. For all sizes, the relative distance between the peaks is maintained, but upon increasing $N$ the high-$q$ decay of $P(q)$ shifts further and further down, approaching the experimental curve. It is important to point out that the observed dependence on size for $P(q)$ is also present in real microgels of different sizes, with the peaks becoming shallower for small microgels. From the estimated values of $\gamma$ for each simulated size, we get an effective size of the monomer bead, amounting to $\approx 4$ nm for the largest microgel. We stress that in order to reach a realistic value of the PNIPAM monomer size $\sigma\sim 1$ nm, we should increase
the number of monomers up to $N \approx 2 \times 10^7$, which is unfeasible with present day computational techniques. Such a discrepancy in size between simulated and experimental microgels thus explains the high-$q$ deviations of the numerical form factors observed in Fig.~\ref{size_effects}(a). In addition, the numerical form factors at large wave-vectors can be well-described by an inverse power law, $P(q)\approx q^{-n}$, with $n\sim 1$ for all investigated cases. The fact that $n$ does not vary with system size suggests that microgels with different $N$ possess the same topological structure, at least on a mesoscopic scale. Finally, we notice that $P(q)$ of the unperturbed microgel also shown in Fig.~\ref{size_effects}(a) presents numerous peaks in agreement with a homogenous dense spherical system, significantly deviating  from the experimental findings  for both the relative position of the peaks and for the shape of the curve at small $q$. 

To better visualize and quantify the differences between the various system sizes, we report in Fig.~\ref{size_effects}(b) the density profiles of the three different systems as well as the corresponding snapshots in Fig.~\ref{size_effects}(d). As expected, the surface contributions are found to dominate for small-sized microgels of a few thousands monomers, while they become less and less relevant when increasing $N$. In all cases, the core behavior is rather similar, while the corona becomes more and more structured only for larger microgels. This result is the  real space counterpart  of the stronger pronunciation of the peaks of $P(q)$ with increasing microgel size.
In the unperturbed case we find that the system is homogeneous and the size of the corona is rather insignificant, as it can be observed in the inset of Fig.~\ref{size_effects}(b) and
in the corresponding snapshot in Fig.~\ref{size_effects}(d). 

Further information on the microgel internal topology is obtained by focusing on the chain length distribution as a function of $N$, which is reported in Fig.~\ref{size_effects}(c).
Defining the chain length $l$ as the sequence of monomers included between two subsequent crosslinkers, we compare its distribution $N_l$ for all the investigated cases.
As shown in a previous work~\cite{Rovigatti2017},
the assembly without a designing force leads to a network structure in good agreement with the Flory theory. Instead, the introduction of the force gives rise to a larger
number of chains with lengths $l>50$.  This effect holds for all studied $N$, although the probability of having longer chains clearly increases with size. While in the absence
of the force $N_l$ is well described by a single exponential, when the force is introduced we have that the same exponential only holds for relatively short chains, while
a second exponential decay is found to describe the distribution for large chain lengths. These results confirm that, in the presence of the force upon changing $N$, the
internal topology of the network is preserved. 

Overall, changing system size, we observe small differences in the density profiles (also due to statistics) and more pronunced ones in the form factors. These are the consequences of the fact that the surface-to-volume contributions play a different role on the final assembled structures. Notwithstanding this, our protocol is now able to generate microgels with a similar topology and core-corona ratio independently of size and we will further show below that, thanks to this, the comparison with experiments does not depend quanlitatively, but only quantitatively, on $N$.
 Consequently, the system size becomes a parameter that can be optimised in order to reproduce the properties of interest while, at the same time, reducing the computational effort.

\subsection{Comparing experiments and simulations}
Having discussed the general effect of the force on the structure of the microgels, we now perform a detailed comparison of the experimental form factors with those calculated for the largest simulated microgels for all studied temperatures. This also allows us to establish a mapping between temperature in $\degree$ C and the solvophobic parameter $\alpha$.

\subsubsection{Form factors and temperature mapping} 
\begin{figure*}[!htb]
  \minipage{0.5\textwidth}
  \includegraphics[width=\linewidth]{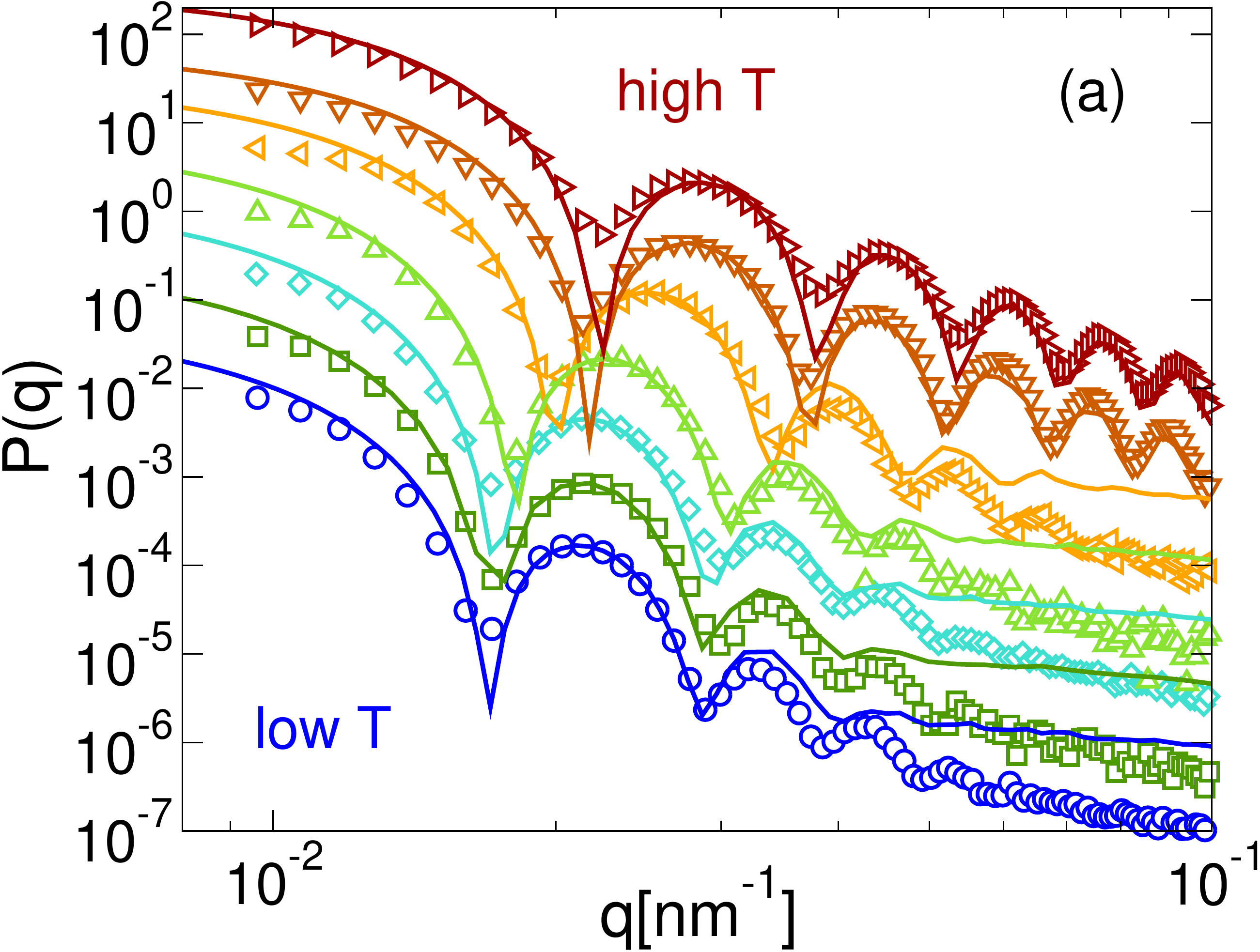}
  \endminipage\hfill
  \minipage{0.5\textwidth}
  \centering
  \includegraphics[width=0.88\linewidth]{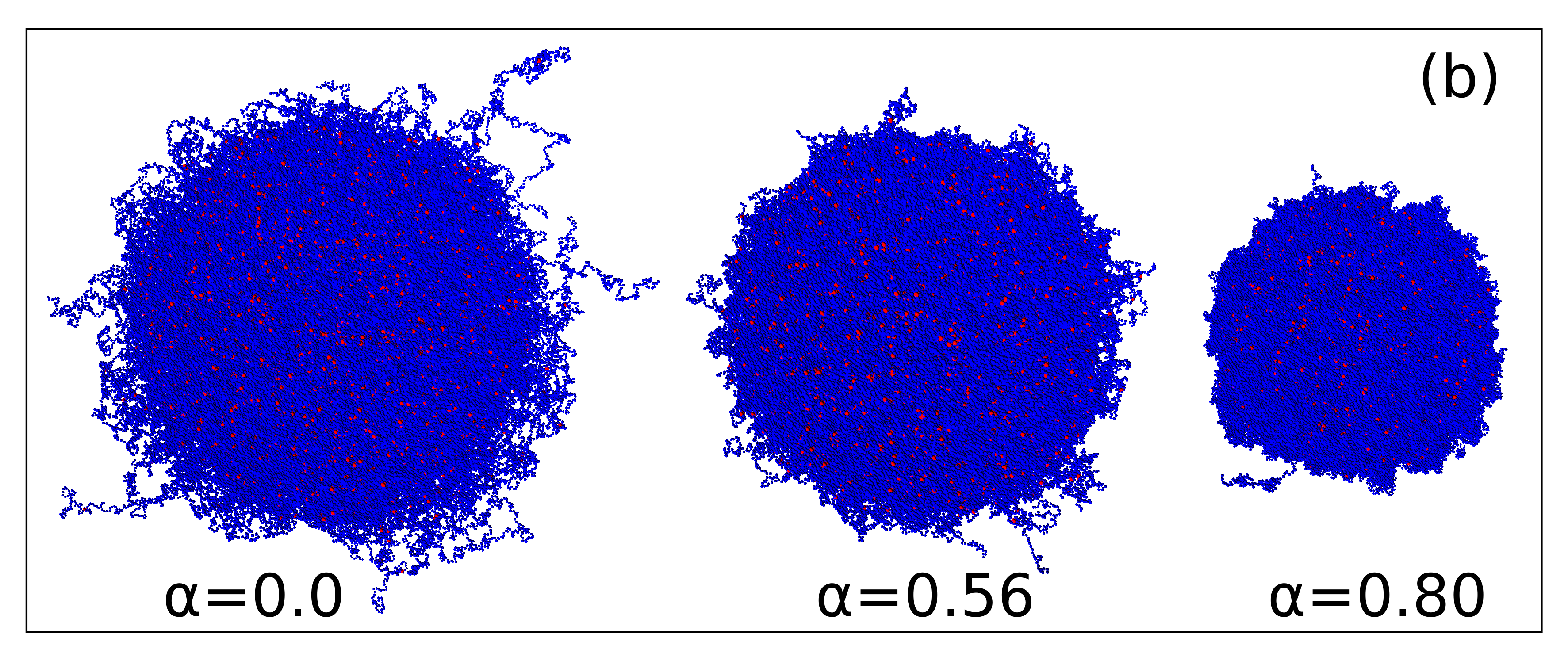}
  \includegraphics[width=0.95\linewidth]{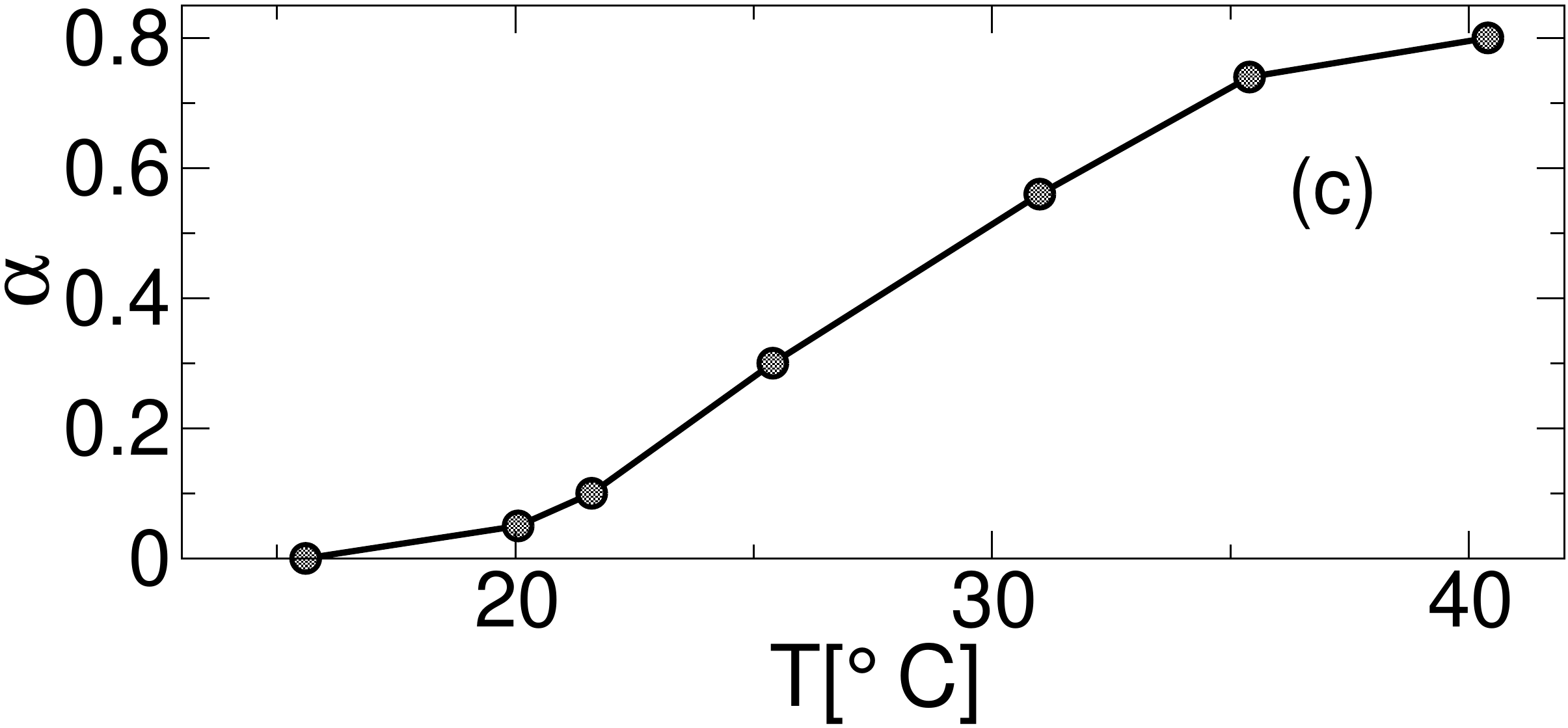}
  \endminipage\hfill
  \caption{ \label{fig2} (a) Comparison between experimental (empty symbols) and numerical (Eq.~\eqref{eq:Pq}, full lines) form factors for $N\approx 336000$. The $x$-axis is rescaled by $\gamma=q^*_{\rm exp}/q^*_{\rm sim}=0.2326$, where $q^*$ is the position of the first peak of $P(q)$. Different colors correspond to different temperatures $T$ and solvophobic parameters $\alpha$, increasing from bottom to top: $T= 15.6$, $20.1$, $21.6$, $25.4$, $31.0$, $35.4$, $40.4~\degree {\rm C}$ in experiments and $\alpha=0.00$, $0.05$, $0.10$, $0.30$, $0.56$, $0.74$, $0.80$ in simulations. Data at different $T,\alpha$ are rescaled on the $y$-axis with respect to the lowest temperature in order to help visualization; (b) snapshots of the $N\approx 336000$ microgels for $\alpha=0.0,0.56,0.80$ with monomers represented in blue and crosslinkers in red; (c) mapping between $T$ and $\alpha$.}
\end{figure*}

We compare the numerical and experimental $P(q)$ for the largest studied microgels ($N\approx336000$) in Fig.~\ref{fig2}(a) for several values of the temperature and corresponding $\alpha$. Up to the second peak, the agreement between experiments and simulations is remarkably good at all $T$. The fact that the numerical data present peaks that are sharper and deeper  could be explained by the presence of a weak polydispersity in the experimental data, that is not considered in the simulations. Most importantly, the positions of all the visible peaks in the simulations are found to coincide with those in experiments.  At high $T$, where the microgel collapses and becomes more homogeneous, the agreement improves even further, with the numerical data being able to capture the positions and heights of all measured peaks. We notice that the deviations occurring
at large $q$ are entirely attributable to the smaller size of the numerical microgels as compared to the laboratory ones, as discussed in the previous section,  leading to a different structure at very short length-scales. Representative snapshots of the microgel across the volume phase transition are shown in Fig.~\ref{fig2}(b). It is evident that the inhomogeneous corona still retains a large degree of roughness even for temperatures above the VPT, differently from what observed in the case of microgels with a more homogeneous structure\cite{Gnan2017}.

We stress that the comparison in Fig.~\ref{fig2}(a) is obtained with the same value of the scaling factor $\gamma$ obtained for $\alpha=0$, that is maintained for all temperatures.
However, we adjust the value of the solvophobic interaction strength $\alpha$ in order to capture the $T$-variation of $P(q)$.  The resulting
relationship between $\alpha$ and $T$ is illustrated in Fig.~\ref{fig2}(c). We find that
an approximately linear dependence holds at intermediate temperatures, showing some deviations at low and high $T$. While the former may be due to the arbitrary
choice of the $\alpha=0$ value with the lowest available $T$, the latter is more likely related to the implicit nature of the solvent employed in the simulations. 
These results also confirm the appropriateness of the $V_{\alpha}$ potential, here tested for the first time against experiments across the VPT.

\begin{figure}[!htb]
 \minipage{0.5\textwidth}
\includegraphics[width=\linewidth]{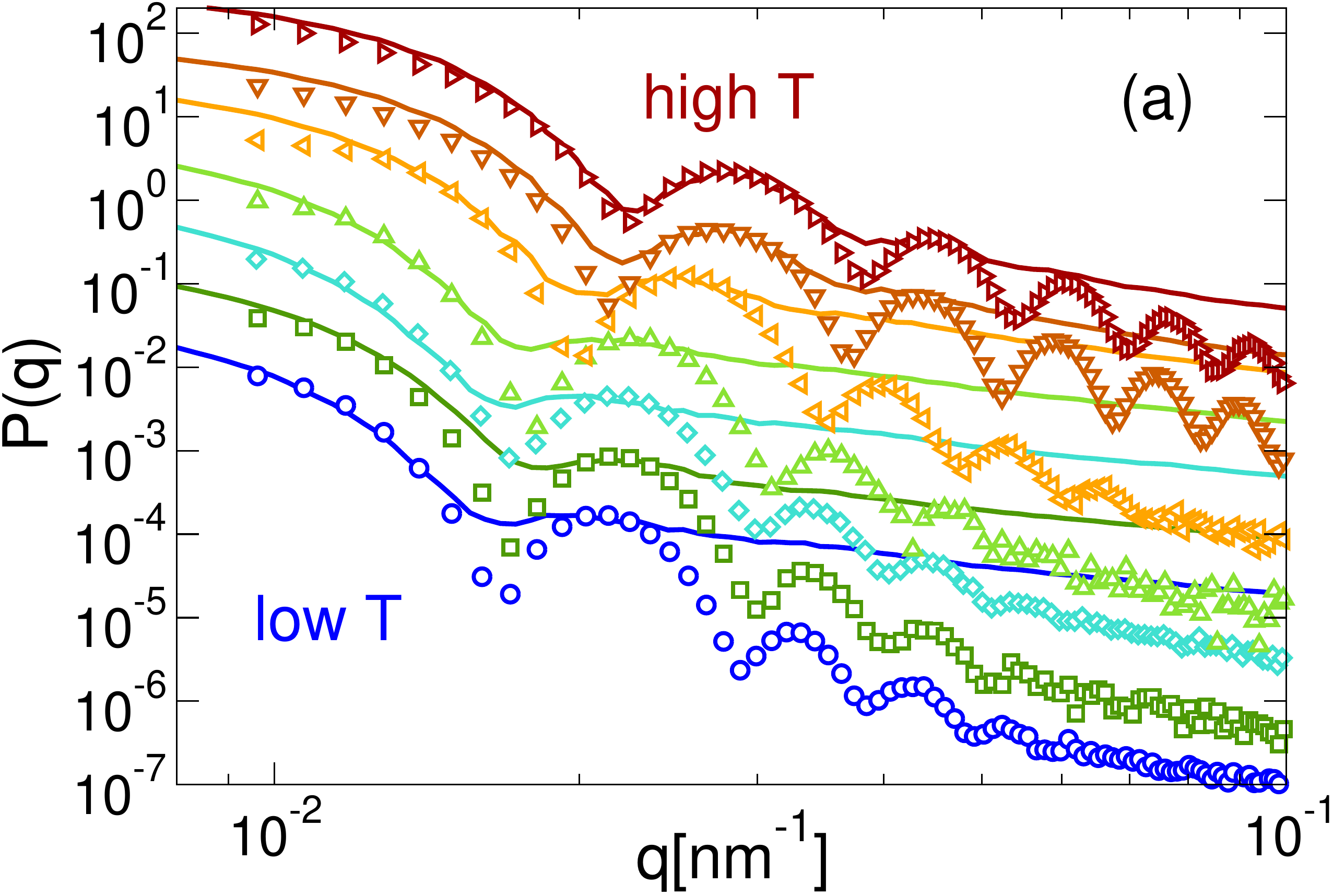}
\endminipage\hfill
\minipage{0.5\textwidth}
 \includegraphics[width=\linewidth]{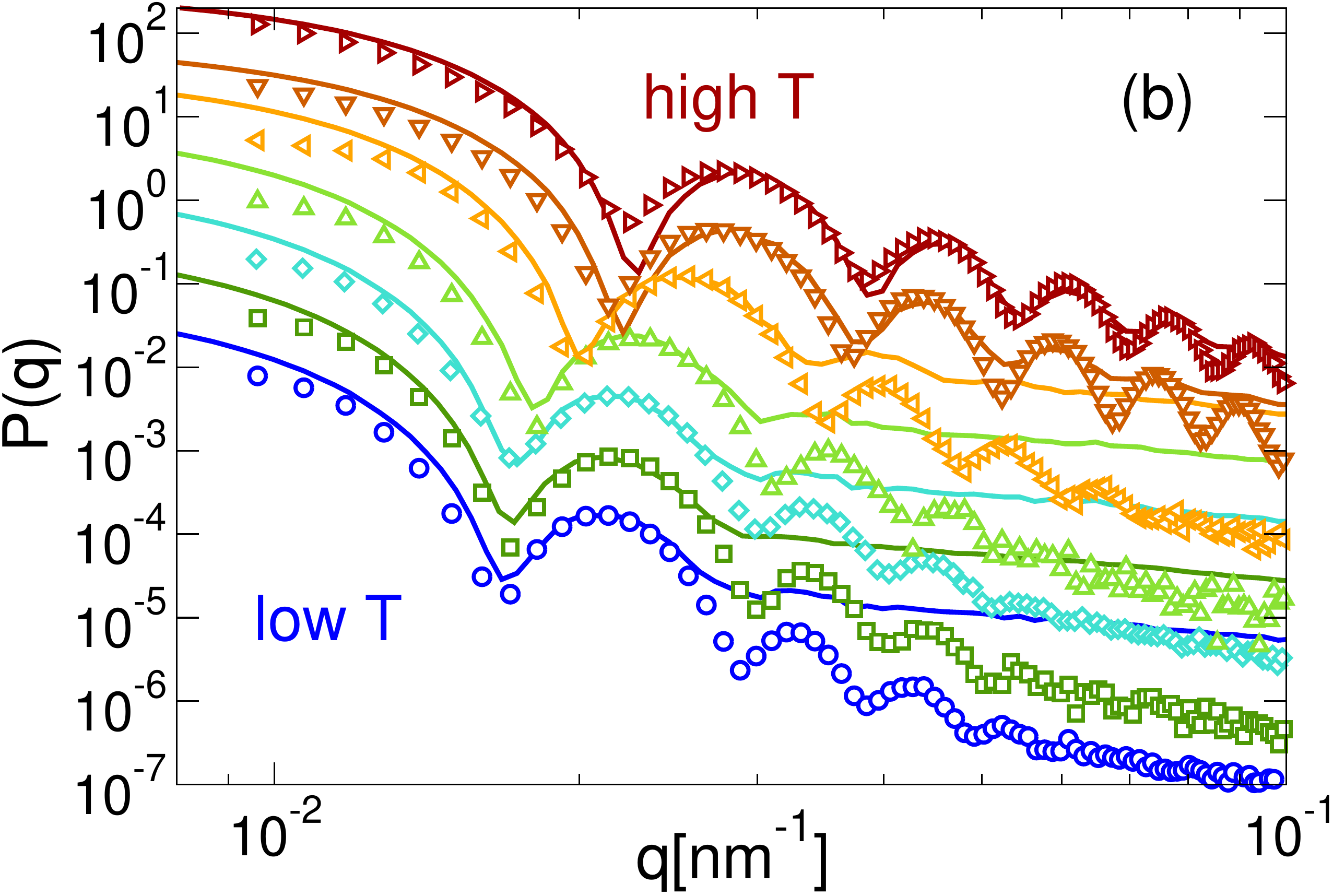}
  \endminipage\hfill
\caption{  \label{fig3}
Comparison between experimental (empty symbols) and numerical (Eq.~\eqref{eq:Pq}, full lines) form factors for (a) $N\approx 5000$ and (b)  $N\approx 42000$. The $x$-axis is rescaled by $\gamma_{5000}=0.0580$ and $\gamma_{42000}=0.124$, respectively. Different colors correspond to different temperatures $T$ and solvophobic parameters $\alpha$, increasing from bottom to top: $T= 15.6$, $20.1$, $21.6$, $25.4$, $31.0$, $35.4$, $40.4~\degree {\rm C}$ in experiments and $\alpha=0.00$, $0.05$, $0.10$, $0.30$, $0.56$, $0.74$, $0.80$ in simulations. These are the same values used for the case $N \approx 336000$. Data at different ($T,\alpha$) are rescaled on the $y$-axis to help visualization.}
\end{figure}

To validate the size independence of our model and the robustness of the ($T,\alpha$) mapping to describe the deswelling transition of the microgels,  we further compare the experimental form factors with those calculated from simulations of different system sizes using the same $\alpha$ values for all $N$.  Again, we keep constant the scaling factors $\gamma$, that we determined for $\alpha=0$.
The comparisons are shown in Fig.~\ref{fig3}(a) and Fig.~\ref{fig3}(b) for the small  and intermediate size systems, respectively.
Strikingly, we find the swelling behavior is well captured for each system size. The peaks are indeed found in the position corresponding to those of the experimental curves, even though they are barely visible, especially for the smallest studied system. The high-$q$ deviations between experiments and simulations become more evident as $N$ decreases, but the agreement improves at high $T$. From these results, we can conclude that the relationship between $T$ and $\alpha$, shown in Fig.~\ref{fig2}(c) is unaffected by size effects. Thus, even though smaller systems give rise to a worse $q$-space resolution, a size-independent swelling behavior is found for all studied $N$, confirming the reliability of our procedure in reproducing experimental results.

\subsubsection{Density profiles and swelling curves} 

\begin{figure}[h]
 \minipage{0.5\textwidth}
  \includegraphics[width=\linewidth]{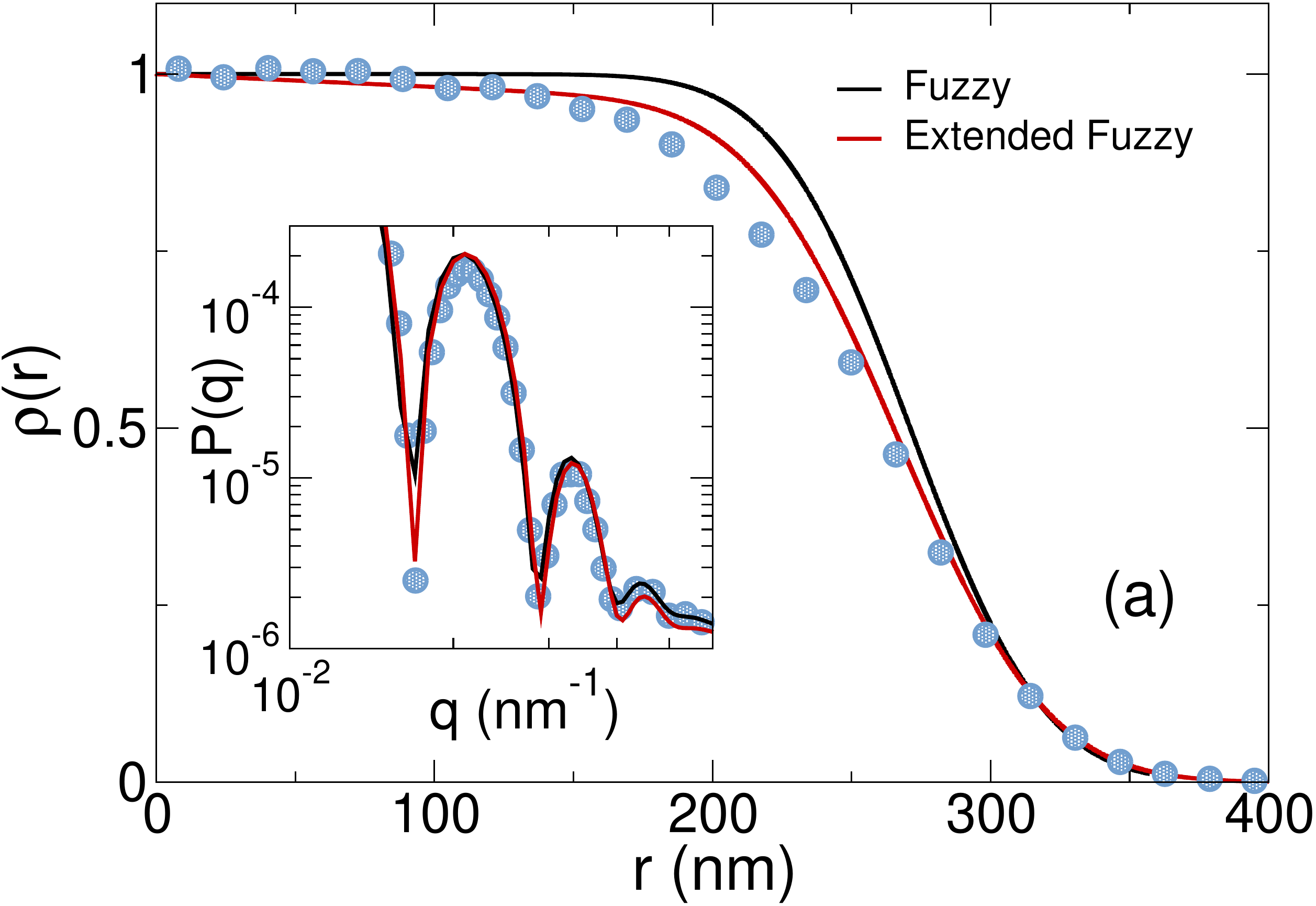}
  \endminipage\hfill
\minipage{0.5\textwidth}
  \includegraphics[width=\linewidth]{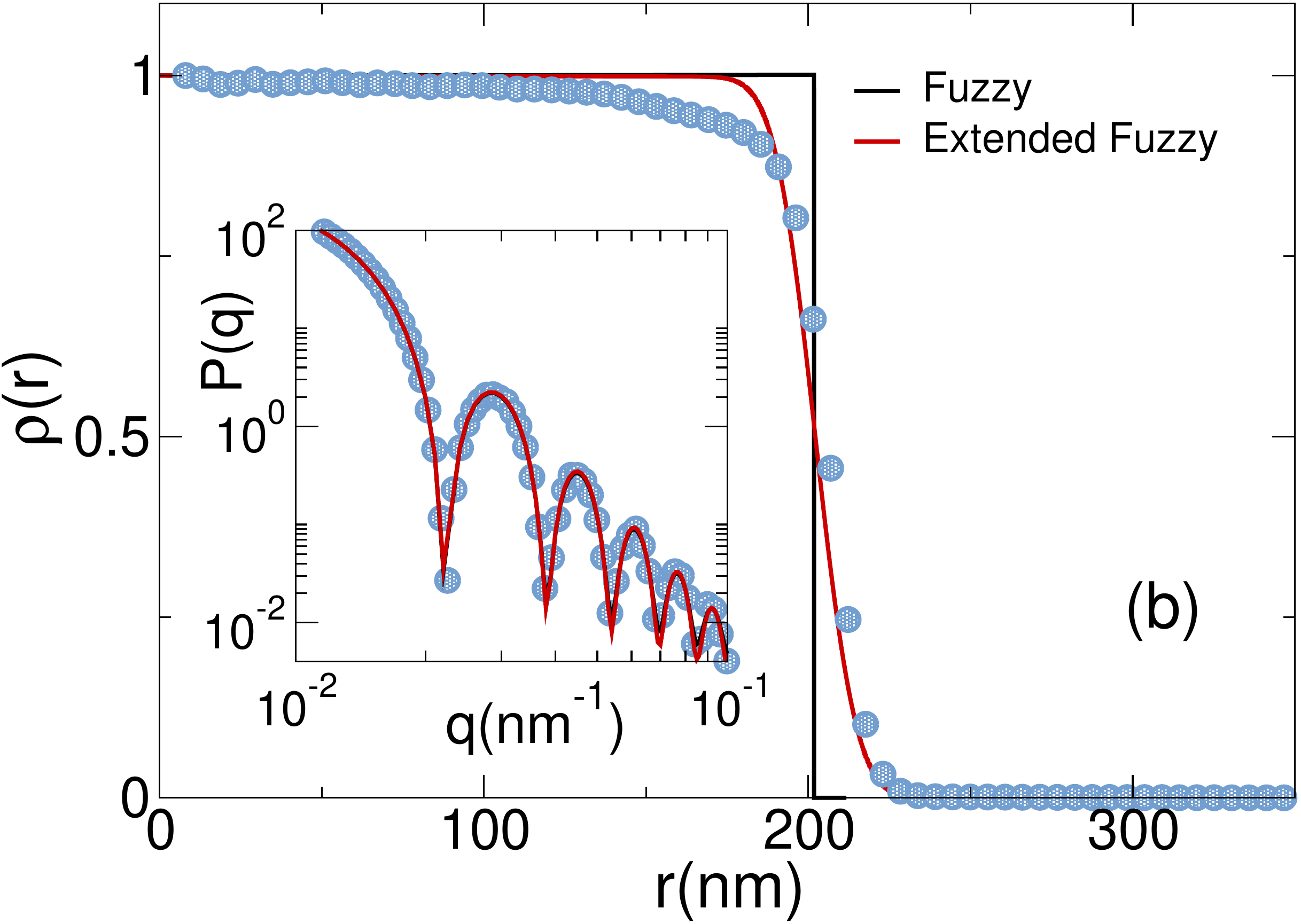}
  \endminipage\hfill
\caption{  \label{figDens} Comparing density profiles for the standard (black) and generalized (red) fuzzy sphere models with the numerical results (blue symbols) for $T=15.4^{\circ}$C (a) and  $T=40.4^{\circ}$C (b). All data are rescaled to 1 at $x=0$ for clarity.}
\end{figure}

In order to directly visualize the internal structure of the microgel, we move to real space. To obtain the experimental radial density distributions from the scattering data, we need to fit the measured form factors. Building on the evidence from recent super-resolution microscopy experiments~\cite{Bergmann2018}, we employ an extended
fuzzy sphere model (Eqs.~\eqref{eq:rho},\eqref{fuzzyX}).
We also calculate $\rho(r)$ directly from simulations and then convert them to real units by rescaling the $x$-axis by $1/\gamma$.
We show the comparison between numerical data and the corresponding ones extracted from the fits in Fig.~\ref{figDens} for two representative temperatures, respectively in the swollen (a) and collapsed (b) regimes. 
We stress that the use of the standard fuzzy sphere model with a homogeneous core to fit the experimental $P(q)$ not only is at odds with super-resolution data~\cite{Bergmann2018}, but also yields density profiles that
are in worse agreement with numerical data, as shown in Fig.~\ref{figDens}.
Indeed, the generalized fuzzy sphere model agrees very well with the numerical data both in the inner part of the core and in the corona. For intermediate values of $r$ there are some small deviations, mainly due to the non-linear decrease of the density profile.
On the other hand, the standard fuzzy sphere model shows a weaker agreement with the calculated profile, especially due to the presence
of the completely homogeneous core. The disagreement becomes more evident at high $T$ where the standard fuzzy sphere results show a step-like behavior. Instead, a continuously decreasing profile is still observed in simulations and for the generalized fuzzy sphere model, again in close agreement to each other.  The quality of the extended fuzzy sphere fits to $P(q)$ is rather good, as shown in the insets of  Fig.~\ref{figDens}(a,b). From the fits, we estimate the linear correction $s$ to be always quite small, $s<5\times 10^{-4} \text{nm}^{-1}$.  Most importantly, we find small but finite values of $\sigma_{\rm surf}$ also above the VPT, which is consistent with the fact that, even in the collapsed state, the microgels still contain a large amount of water~\cite{Bischofberger2015}. 
This is confirmed by the snapshots of the collapsed state shown in Fig.~\ref{fig2}(b), where many dangling ends are clearly visible in the swollen state, giving rise to a rough profile of the microgel also in the collapsed state.
   
\begin{figure}[h]
\minipage{0.5\textwidth}
  \includegraphics[width=\linewidth]{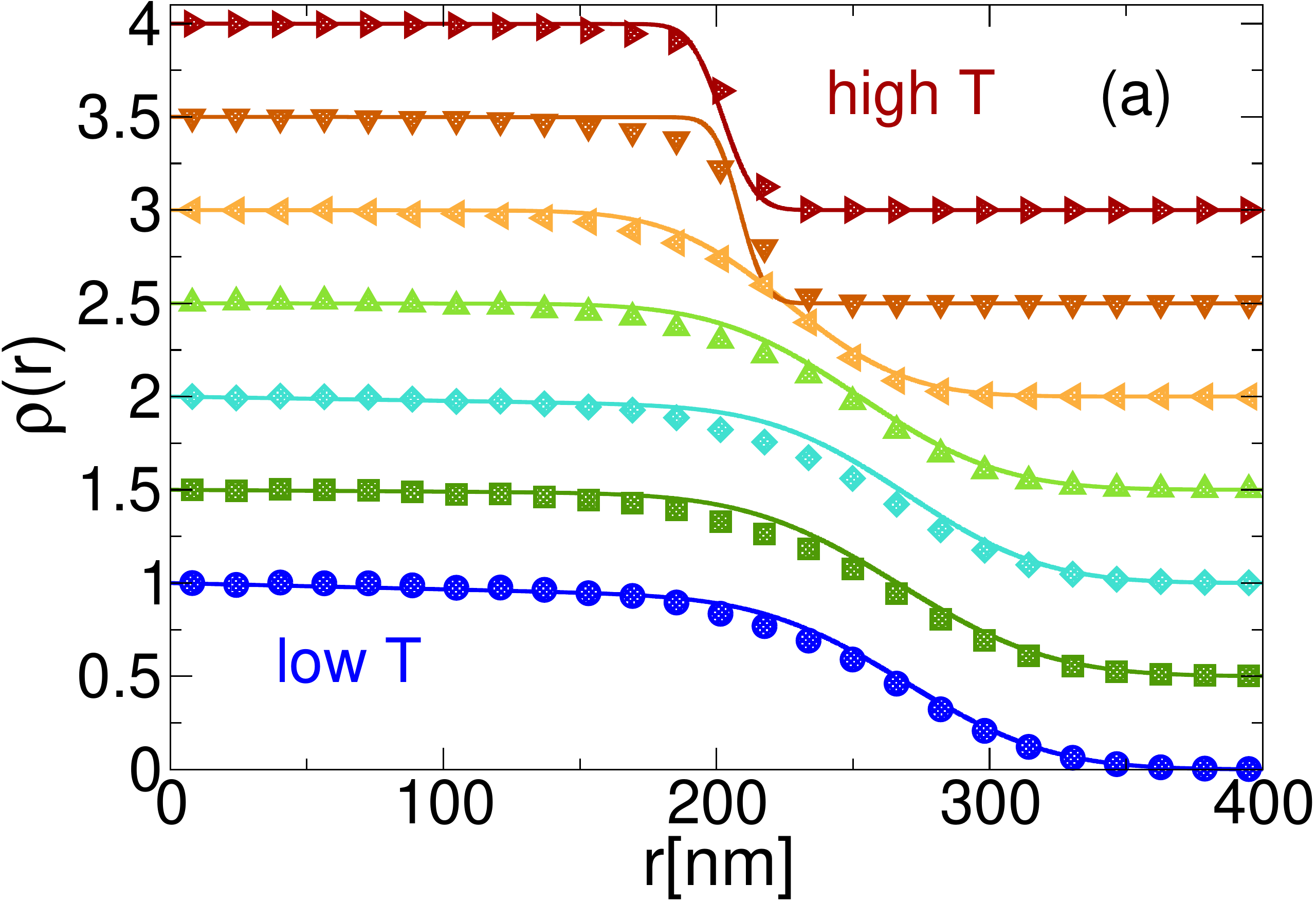}
   \endminipage\hfill
\minipage{0.5\textwidth}
  \includegraphics[width=\linewidth]{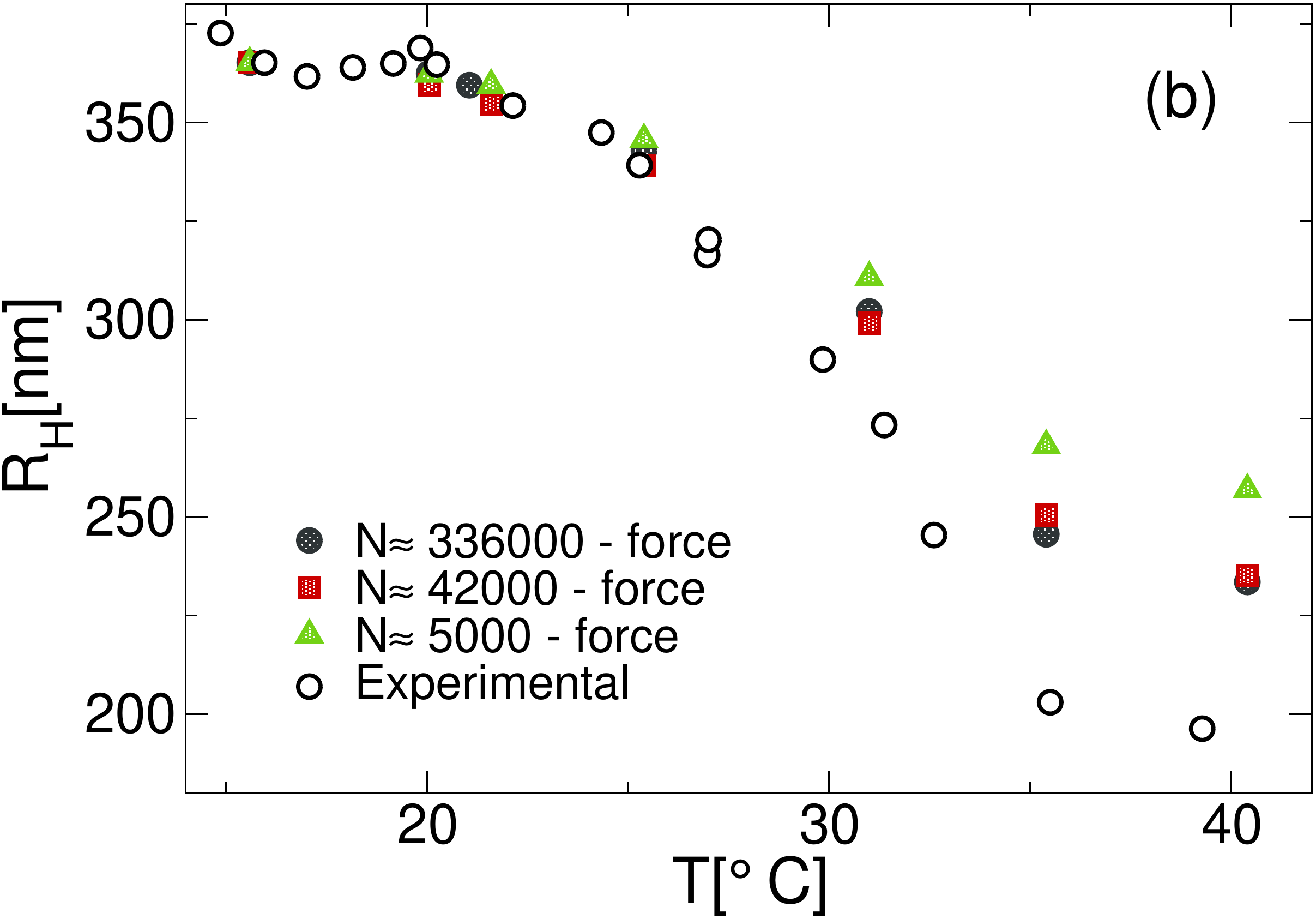}
  \endminipage\hfill
  \caption{ \label{fig:dens2} (a) Comparison between numerical (full symbols) and experimental (full lines) density profiles $\rho(r)$, where the latter are obtained by fitting the form factors to the generalized fuzzy sphere model (Eq.~\eqref{eq:rho}). The numerical $x$-axis is rescaled by $\gamma^{-1}$, while data are normalized to $1$ at the center of the microgel and shifted vertically by 0.5 at different $T$ to improve readability; (b) hydrodynamic radius $R_H$ from experiments (black circles) and simulations (see text for its definition) for $N\approx 336000$ (blue squares), $N\approx 42000$ (red diamonds), $N\approx 5000$ (green triangles), respectively. Numerical data are rescaled to match experiments at low $T$. }
\end{figure}
A summarizing comparison between numerical and experimental density profiles is reported in Fig.~\ref{fig:dens2}(a) for all studied temperatures. The agreement is again found to be very good throughout the whole $T$ range for both the core and the corona regions. 

Finally, we discuss the comparison of numerical and experimental results through the swelling curve, rather than through the form factors and/or the density profiles. This amounts to comparing the measured hydrodynamic radius $R_H$, measured in Dynamic Light Scattering experiments on dilute samples, with its numerical analogue. However, the latter is not unambiguously determined in our simulations, and expensive treatments including hydrodynamic interactions would be needed for a correct assessment of its value. To circumvent this problem, we adopt an operative definition for $R_H$, which is assumed, as in previous works\cite{Gnan2017}, to be 
the distance at which $\rho(R_H)=10^{-3}$. This choice, when converted with the determined  $\gamma$ factor, provides a numerical estimate of $R_H$ in good agreement with experiments for the swollen state. Experimental and numerical $R_H$ are reported in Fig.~\ref{figDens}(c) for the various investigated system sizes. A sharp change of $R_H$ is observed with increasing temperature both in experiments and in simulations, but for the latter we find that the collapse is less
pronounced than in experiments and it further reduces with decreasing system size. This may be due to steric effects of the bead size in the collapsed state, which are more important for small microgels. However, the dependence of $R_H$ to $N$ seems to saturate at the largest investigated sizes, so that the discrepancy between simulations and experiments seems to remain present even by extrapolating $N$ to realistic values ($\sim O(10^7)$). A possible explanation of this result could be attributed to charge effects, which have been shown to become relevant above the VPT temperature~\cite{Truzzolillo2018,delmonte2019numerical}, and therefore should be taken into account for a more faithful representation of the behavior of the numerical $R_H$ at high $T$. A second possibility would be a fine tuning of the interaction potential between the beads beyond $V_{\alpha}$ in order to obtain a polymer chain elasticity in closer agreement with the experimental one. Such a study is currently in progress.

\section{Conclusions}

In this work we have shown how computer simulations 
can realistically model thermoresponsive microgels by adopting a designing force during the network assembly which can be tuned to quantitatively reproduce experimental form factors for a wide range of temperatures across the VPT. Even if the protocol itself is not meant to reproduce the experimental conditions of the synthesis, it is nevertheless able to generate networks with topologies that can closely match experimental data. 
We have shown that our method is robust to system size for swelling properties and it reproduces very well the experimental form factors, with the agreement improving with the size of the microgels. It is worthwhile to note that the comparison is good even for microgels composed of only a few thousands monomers, which can be routinely studied in simulations. This allows us to establish a relationship between the solvophobic strength $\alpha$ used in simulations and the experimental temperature $T$, finding that they are linearly related across the VPT, as expected. Such a relation is found to hold for all studied system sizes. 
Our results open up the possibility to address numerous questions about microgels, both at the fundamental and application level. For examples, it has already been successfully applied to the case of microgels at liquid-liquid interfaces, favourably comparing with experimental results as a function of different crosslinker concentrations\cite{camerin2019microgels}.
In addition, the protocol developed here is very general, making it possible to use it to tackle the investigation of microgels with different density profiles, such as homogeneous\cite{Mueller2018}, hollow~\cite{Nayak2005,Scotti2018} and ultrasoft ones~\cite{Gao2003,Bachman2015,Virtanen2016}, nowadays synthesized in experiments.
Furthermore, we plan to use the knowledge gained at the single-particle level to tailor the materials properties of bulk systems in the near future.

\section{Acknowledgments}
We thank M. Paciolla for early contributions to this work. AN, FC, NG, LR and EZ acknowledge financial support from the European Research Council (ERC Consolidator Grant 681597, MIMIC). PS acknowledges financial support from the European Research Council (ERC-339678-COMPASS) and the Swedish Research Council (VR 2015-05426). 
JJC thanks the German Research Foundation (Collaborative Research Center SFB985).
SAXS experiments were performed at the cSAXS beam line of the Swiss Light Source SLS at Paul Scherrer Institute, and we gratefully acknowledge the help of the local contact A. Menzel.

\bibliography{draft}

\end{document}